# Fermi Level Depinning in Two-Dimensional Materials *Using a* Fluorinated Bilayer Graphene Barrier

*Cunzhi Sun[§,†,‡], Cheng Xiang[†,‡], Rongdun Hong[§], Feng Zhang[§], Timothy J. Booth[†,‡], Peter Bøggild[†,‡,]\*, and Manh-Ha Doan[†,‡,]\**

[§] Department of Physics, Xiamen University, Xiamen, 361005, P. R. China

[†] Department of Physics, Technical University of Denmark, Kgs. Lyngby 2800, Denmark

[‡] Centre for Nanostructured Graphene (CNG), Technical University of Denmark, Kgs. Lyngby 2800, Denmark

**ABSTRACT:** Strong Fermi level pinning (FLP) - often attributed to metal-induced gap states at the interfacial contacts - severely reduces the tunability of the Schottky barrier height of the junction and limits applications of the 2D materials in electronics and optoelectronics. Here, we show that fluorinated bilayer graphene (FBLG) can be used as a barrier to effectively prevent FLP at metal/2D materials interfaces. FLBG can be produced via short exposure (1-3 min) to $SF_6$ plasma that fluorinates only the top layer of a bilayer graphene with covalent C-F bonding, while the bottom layer remains intrinsic, resulting in a band gap opening of about 75 meV. Inserting FBLG between the metallic contacts and a layer of $MoS_2$ reduces the Schottky barrier height dramatically for the low-work function metals (313 and 260 meV for Ti and Cr, respectively) while it increases for the high-work function one ( 160 meV for Pd), corresponding to an improved pinning factor. Our results provide a straightforward method to generate atomically thin dielectrics with applications not only for depinning the Fermi level at metal/transition metal dichalcogenide (TMD) interfaces but also for solving many other problems in electronics and optoelectronics.

**INTRODUCTION**

As the number of transistors per CPU has increased exponentially in the last decades, a trend known as Moore´s law, various challenges are emerging in the conventional silicon semiconductor technology when the transistor is scaled down to sub-10 nanometers[1,2]. Two-dimensional (2D) materials, such as graphene and transition-metal dichalcogenides (TMDs), have drawn tremendous attention recently due to their promising advantages including atomic thickness and smooth channel-to-dielectric interface without dangling bonds[3,4]. However, while the absence of bandgap in graphene results in a high off-current of the device, a high contact resistance of metal/TMD junctions makes it difficult to inject charge carriers to the channel[5,6]. Opening a bandgap of graphene and reducing the contact resistance of TMDs, therefore, are the important directions to bring these materials to pragmatic switching applications.

Strong Fermi level pinning (FLP) is generally observed in metallic contacts to the 2D materials, originating from metal-induced gap states at the interface[7-9]. This renders the Schottky barrier height (SBH) of the junction uncontrollable and leads to increase of the contact resistance[10]. Traditionally, doping semiconductors at the contact region and/or using a metal-insulator-semiconductor structure can release the FLP and allow for an ohmic contact[11,12]. Since strong, highly local doping of 2D materials is not yet a practical route to solving the problem, insertion of a buffer layer between the metal contact and TMDs is considered an effective alternative way to alleviate the FLP[8-10]. Indeed, by insertion of a thin insulating layer, such as $TiO_2$, ZnO, or h-BN at the metal/$MoS_2$ interfaces, the SBH can be significantly reduced, improving the pinning factor and the contact resistance[13-15]. The thickness of the insulator should be carefully controlled, since a thick layer would block the FLP but also increase the tunneling resistance. Using graphene as an interlayer could help to reduce the SBH at the metal/TMD interface but does not alleviate the FLP because graphene itself is a (semi)metal. It has been reported that strong fluorinating agents, namely xenon difluoride ($XeF_2$), can turn monolayer graphene to an insulator with a band gap of 3 eV due to formation of $sp^2$ C-F bonding on both sides, creating fluorographene[16]. However, stacking the double-side fluorinated graphene monolayer on a TMD will not result in smooth and stable contacts due to the covalent bonding of F atoms at the graphene/TMD interface.

In this study, we show that fluorinated bilayer graphene (FBLG) can be used as an interlayer to effectively prevent FLP at metal-2D material interfaces. Raman scattering and X-ray photoelectron spectroscopy (XPS) measurements reveal that after 1-3 minutes of exposure to $SF_6$ plasma only the top layer of bilayer graphene is fluorinated with covalent C-F bonds, while the bottom layer remains intrinsic and free of dangling bonds and F species. Fluorination of mono- and bilayer graphene has previously been observed to lead to opening of a bandgap in the BLG[17]. As a proof of concept, we fabricate and investigate the transport properties of metal/FBLG-contacted $MoS_2$ transistors, using contact metals with different work functions. The extracted SBH in control devices without a FBLG layer at the metal/$MoS_2$ junctions is almost identical for all metals as characteristic for strong FLP. On insertion of FBLG, the SBH dramatically reduces for the low-work function metals, resulting in higher on-current of the transistor. In addition, it alleviates the line-up effect of the work function of the metal at the interface, leading to a better pinning factor. Our results define a straightforward method to generate thin dielectrics for depinning the Fermi level at metal/TMD interfaces in 2D material-based devices, as well as for various applications in advanced electronics and optoelectronics, and can be readily applied at the wafer scale using chemical vapor deposited graphene.

**RESULTS AND DISCUSSION**

To demonstrate a possible gap opening in FBLG, we fabricate field-effect transistors (FET) using exfoliated bilayer graphene on $SiO_2$/Si substrates both with and without applying the fluorination process, and investigate their electrical transport properties (Figure 1). BLG flakes were identified from the visual contrast in an optical microscope[18] and verified by Raman scattering and atomic force microscopy (AFM) measurements. The samples were annealed at 300 ºC in Ar/$H_2$ environment for 2 hours to remove residues on their surfaces. Fluorination is carried out in an inductively coupled plasma (ICP) etcher with $SF_6$ precursor gas under a high vacuum of $10^{-6}$ Torr over a typically short time (1-2 min) in attempt to limit fluorination to the top layer. To verify this, we used Raman spectroscopy and X-ray photoelectron spectroscopy (XPS) to examine the sample after fluorination. Figure 2a shows the evolution in the Raman spectrum of BLG with increasing fluorination time. Features corresponding to defect active bands are observed during fluorination, which include a blue-shift of the G peak, reduced intensity and broadened full width of half-maximum (FWHM) of the 2D peak, and the appearance of the D and D´ peaks. Figure 2b shows the trends of the 2D peak as a function of fluorination time. The 2D peak of pristine BLG is well fitted by four Lorentzian components while after 185 seconds of fluorination it is best fitted with only one component (see SI, Figure S1). This is attributed to a progressive transform of the 2D band of the bilayer into one of a monolayer. The enlarged FWHM of the 2D peak is due to loss of the AB-stacking configuration caused by induced-defects on the top graphene layer[16,19,20]. It is also noted that the FWHM of the D peak in our fluorinated BLG is about 40 cm$^{-1}$ (see SI, Figure S1), which is in line with the previous studies on the Raman spectra of single-side functionalized BLG[19]. This indicates that only the top layer of BLG is fluorinated, while the bottom layer remains pristine, with C-F covalent bonds forming only on surface of the top layer of BLG, as illustrated Figure 1a.

Metal contacts of Cr/Au (20/80 nm) are made on pristine and fluorinated graphene using standard electron-beam lithography and e-beam evaporation[21]. The current-voltage characteristics of the BLG device shows a linear behavior with high conductivity, while it is nonlinear with much lower current in the FBLG device, as seen in Figure 1c. Using Si as a back-gate through the SiO$_2$ insulating layer, we clearly observe the distinctive features in the transfer curve of FBLG compared to that of pristine BLG including i) increased resistance (without, however, reaching the insulating regime as in the case of fluorographene[16] or perfluorographene[20]), ii) improved on-off ratio, and iii) a large 40 V shift of the charge-neutrality point towards the positive gate voltage. We attribute these effects to a small bandgap opening and p-doping in BLG after fluorination[22-24].

The fluorination induces an asymmetry between the top and bottom graphene layers due to differences in carrier concentrations and/or a disordered lattice structure. Varying the back-gate voltage increases this inequivalence further. As a result, changes in the Coulomb potential lead to opening of a gap between valence and conduction bands of BLG[24]. To estimate the electrical bandgap of FBLG, we adopt a well-known method built by Xia et al., which assumes that the off-current of the BLG transistor is proportional to $\exp(-q\phi_{\text{barrier}}/k_\text{B}T)$, where $q$ is the electron charge, $\phi_{\text{barrier}}$ is the SBH at the interface of BLG and metal electrodes, $k_\text{B}$ is the Boltzmann constant, and T is the temperature[25]. The bandgap of BLG is double the size of the SBH at the charge neutrality point. Then, the increased bandgap of fluorinated BLG is as $\Delta E_\text{g} = 2\Delta(\phi_{\text{barrier}}) = 2(k_\text{B}T/q)\ln(I_{\text{off}}^0/I_{\text{off}})$, where $I_{\text{off}}^0$ and $I_{\text{off}}$ are the off-currents of pristine and fluorinated graphene, respectively. The bandgap of our FBLG is estimated to be about 75 meV, which is in a similar range of the molecular doped BLG reported in references 22 and 23.

Figure 3 presents the XPS spectra of BLG after 185 sec fluorination. In the C1s spectrum region (Figure 3a), beside the strong and sharp peak of the C-C sp2 bond centered at about 284 eV, we clearly observe the satellite peaks relating to the C-F bonds. The peak (red color) at about 285.2 eV correspond to the C-CF sp3 covalent bonds[26]. The features at around (violet) 287, (blue) 290.7 (blue) and (green) 292 eV are attributed to the $C-F_{2,3}$, $C-F_2$ and $C-F_3$ groups, respectively[26,27]. The relative amount of these components extracted from the fitting procedure is about 30%, 7.6%, 34%, and 3.6%, respectively. It should be noted that these amounts may not represent the exact fractions of the chemical groups on the sample due to the complexity of the surface chemistry of FBLG. This does, however, confirm the existence of the C-F bonds on the surface of BLG after fluorination. Figure 3b shows the XPS spectrum of the FBLG in the F1s region, confirming the co-existence of the covalent and so-called semi-ionic F-C bonds on FBLG[27].

Having established that the BLG is transformed into FBLG with a small bandgap, we now use it as an interlayer between metal contacts and semiconducting TMDs to investigate the Fermi level pinning effect. We choose $MoS_2$ as the conducting channel for the fabricated FET device, since this material is most intensively studied among TMDs with promising advantages for various applications in field-gated electronics[28], optoelectronics[29], spintronics[30], valleytronics[31], sensors[32] and so on[33]. Multilayer $MoS_2$ with a thickness of about 10 nm is first mechanically exfoliated on a $SiO_2$/Si substrate using the scotch-tape method[34]. BLG is produced by the mechanical exfoliation method and stacked on the $MoS_2$ flake by aligned transfer under a microscope[34,35]. The sample then is coated by a polymethyl methacrylate (PMMA) layer, and the contact regions are opened by using the electron-beam lithography process. Fluorination is applied to induce FBLG, and metals with different work functions including Ti, Cr, Au, and Pd with a thickness of 20 nm are deposited as the electrical contacts. To protect the metals from oxidation, all the contacts are covered by a thick Au layer (100 nm). The channel region is defined by another lithography step, and the remaining excess BLG on $MoS_2$ is etched away using oxygen plasma, resulting in the FET structure shown in the inset of Figure 4a. Control devices, wherein the metal contacts are directly deposited on $MoS_2$ or using BLG without fluorination as the insertion layer, are fabricated under the same conditions to allow for a direct comparison. Figure 4 shows the gate-modulated currents (transfer curves) of the $MoS_2$ FET without and with the FBLG insertion layer. It is clearly observed that for low work function metals (Ti and Cr), much higher on-currents in all gate voltage regimes are achieved with insertion of the FBLG layer, while for high-work function metals (Au and Pd), there are crossover points in the transfer curves of the devices without and with FBLG. The increased current of the devices with Ti/FBLG and Cr/FBLG contacts is a result of the reduced

Schottky barrier height at the metal/MoS$_2$ junctions, and the crossover point is signature of the injection of holes into the valence band of MoS$_2$ [36].

To get the effective SBH at the metal contacts and MoS$_2$ interfaces, we adopt the 2D thermionic emission equation to extract the activation energy from the slope of an Arrhenius plot. At a certain temperature, current flow in the channel as[7,9]

$$I_{2D} = W A^*_{2D} T^{3/2} \exp\left(\frac{q\phi_B}{k_B T}\right) \exp\left(\frac{qV_D}{k_B T}\right), \qquad (1.1)$$

where W is the channel width, $A^*_{2D}$ is the 2D equivalent Richardson constant, $\phi_B$ is the SBH, and $V_D$ is the applied drain voltage. We measured transfer curves of the devices at different temperatures, and extracted SBH at each gate voltage by linearly fitting of the plot of ln(I/T$^{3/2}$) versus 1000/T. The extracted SBH, $\phi_B$, is then plotted as a function of gate voltage for all devices as shown in Figure 5a-d. We applied V$_D$ = 1 V to extract the SBH in all devices using comparable conditions. The use of a relatively high V$_D$ contributes to reduce the impact of the Schottky barrier at the drain side[7]. The effective SBH at the contact junction is defined as the activation energy at the flat-band gate voltage, which is seen as the point where $\phi_B$ ceases to decrease linearly with gate voltage[37]. It can be seen in Figure 5a-d that when the metals form direct contact to MoS$_2$, the SBH is just slightly changed (50 meV) from Ti to Pd, even though the difference of work function between these two metals is about 1 eV. This is clear evidence of the strong FLP effect at the metal/MoS$_2$ junctions. By insertion of FBLG, however, the SBH significantly reduces for the Ti and Cr electrodes, consistent with the increased current in the transfer curve above (Figure 4a, b). The reduction of the SBH in case of Au contact is small (Figure 5c), and interestingly, the SBH increases when FBLG is inserted between Pd and MoS$_2$ (Figure 5d).

Experimentally, strong Fermi level pinning observed in 2D material-based devices is usually attributed to several mechanisms including i) penetration of the wave function of metal contact electrons into the 2D crystal, which induces a large metal induced gap states (MIGS) even for an ideal 2D surface without defects[9,10], ii) intrinsic defects and/or impurities in 2D materials, and iii) extrinsic disorder-induced gap states (DIGS) forming during device processing[7-10]. Despite the surface of the TMD materials having no dangling bonds, a high density of intrinsic defects ($10^{10}$ to $10^{11}$ per cm$^{-2}$) is generally observed in thin flakes exfoliated from bulk materials due to presence of impurities, S-vacancies, and/or metal-like defects in the crystal[38,39]. Moreover, deposition of the contacts using an electron-beam evaporation process can introduce defects in the crystal structure of the TMD monolayers or topmost TMD layers, generating the extrinsic disorder-induced gap states[40]. We anticipate that all these factors contribute to Fermi level pinning in our $MoS_2$ devices, although which mechanism is dominant is not clear or easy to determine. As a result, a large SBH forms at the interface, despite that the work function of Ti (4.3 eV) and Cr (4.5 eV) is above or aligned with the conduction band minimum of $MoS_2$. In addition, due to the strong FLP, the Fermi levels of the metals are lined-up and pinned to the MIGS, as illustrated in the inset of Figure 5f (left panel). Insertion of FBLG weakens the FLP owing to the two effects. It prevents the formation of MIGS since distance between the metal contact and $MoS_2$ surface is increased. In addition, FBLG acts as a protection layer to reduce the DIGS formed on the surface of $MoS_2$ in the contact region during metal deposition. These, consequently, reduces the band bending and removes the lining-up of the work functions of the metals, depicted in the inset of Figure 5f (right panel). Insertion of FBLG weakens the FLP owing to the two effects. It prevent the formation of MIGS since distance between the metal contact and $MoS_2$ surface is increased. In addition, FBLG acts as a protection layer to reduce the DIGS formed on the surface of MoS2 in the contact region during metal deposition. These, consequently, alleviates the band bending and removes the line-up of the work function of the metal, depicted in the inset of Figure 5f (right panel). This explains the observation of the opposite trends in SBH of the contacts with low (Ti, Cr) and high (Pd) work functions, after FBLG insertion.

Ideally the SBH, $\phi_B$, is given by the difference between the work function of a metal, $\phi_M$, and the electron affinity of the semiconductor, $\chi$, following the Schottky-Mott (SM) rule[41]

$$\phi_B = \phi_M - \chi \qquad (1.2)$$

However, the MIGS formed at the interface produce deviations from the SM rule and introduces a Fermi level pinning effect. The SBH is then characterized quantitatively by introducing a pinning factor, S, and charge neutrality level, $\phi_{CNL}$, as[41]

$$\phi_B = S(\phi_M - \phi_{CNL}) + (\phi_{CNL} - \chi) \qquad (1.3)$$

The pinning factor is used to estimate how severely FLP occurs at the contacts. In the ideal case, S = 1 (the SM rule) where no FLP at the interface, while zero value of S represents complete pinning. Experimentally, S can be defined as the slope of $\phi_B$ vs. $\phi_M$. Figure 5e shows the S values extracted from the $MoS_2$ devices with the different metal contacts without and with the FBLG interlayer. It is obvious that S is typically small (0.07) when the metals directly contact $MoS_2$, reconfirming strong FLP. We observe that insertion of pristine BLG makes FLP slightly more pronounced. The exact mechanism why pristine BLG leads to a smaller value of S is not clear at this point. Using fluorinated BLG as the interlayer, we would improve the pinning factor to more than 5 times to reach S = 0.37 (the red line). This value is even higher than those reported to date using high-quality thin insulating interface layers such as $TiO_2$, ZnO, or h-BN[13-15]. Figure 5f summarizes our observations of FLP in our $MoS_2$ devices using the different metal contacts without and with FBLG. Insertion of FBLG between the metal and $MoS_2$ results in a significant reduction in available MIGS, lowers the band banding of $MoS_2$, alleviates the line-up of metal work function, and de-pins the Fermi level.

The pinning factor of our MoS2 device with FBLG is about 0.37, which is still not close to the ideal case (S = 1), indicating that there some degree of Fermi level pinning is still remaining at the metal/FBLG /MoS2 interface. This may be because the thickness and/or bandgap of FBLG is not sufficiently optimized. Further studies are being carried out to achieve even better results using CVD-grown graphene, where multiple transfers can be used to optimize the thickness in a practical and scalable manner. Since the fluorination anyway decouples the top-layer from the bottom layer, whether the BLG was initially Bernal stacked or not, the stacking order of the BLG is very likely not to have any influence on the result. This opens for the intriguing possibility of replacing BLG, which is still challenging to grow on a large scale, with two large, sequentially transferred single-layer graphene sheets grown by chemical vapor deposition. Wafer-scale growth and transfer techniques for graphene and other 2D materials have been developed rapidly in the recent years[42-45] and are today mature routes for mass-production. Further study is needed to ultimately prove that non-Bernal stacked CVD graphene bilayers can indeed be used for large-scale de-pinning of the Fermi level of TMD devices and employed in mass production of 2D electronics.

**CONCLUSION**

In summary, our work demonstrates a simple and fast process to convert (semi)metallic BLG to the fluorinated bilayer graphene system where C-F bonds form only on the surface of the top graphene layer. This allows us to open a small bandgap of the top layer while keeping the bottom layer intrinsic and free of dangling bonds. We prove that this FBLG layer can be used as an interlayer between the metal contacts and $MoS_2$ to effectively prevent FLP, and thus reduce contact resistance. Our result provides a new strategy to fabricate new synthetic 2D dielectrics that have the potential to be scaled effectively by replacing the BLG with double-transferred CVD-grown single layer graphene. This atomically thin layer has useful applications not only in de-pinning the Fermi level at metal/TMDs junctions in 2D materials devices, but also paves the way to various applications such as insulating protective coatings for monolayer graphene, dielectric engineering[46] and advanced quantum electronics[47].

## METHODS

**Fluorination of BLG.** BLG was exfoliated from graphite (NGS Naturgraphit, Germany) on $SiO_2$/Si substrate. The bilayer flakes were defined from the visual contrast in an optical microscope, and verified by Raman scattering. The samples were annealed at 300 ºC in Ar/$H_2$ environment for 2 hours to remove residues on their surfaces. Fluorination is carried out in an inductively coupled plasma etcher (ICP, SPTS Technology) using $SF_6$ precursor gas with a platen power of 20W at room temperature under a high vacuum of $10^{-6}$ Torr. The fluorination time was typically short (1-2 min) in attempt to primarily affect the top layer.

**Fabrication of BLG and FBLG device.** Metal contacts of Cr/Au (20/80 nm) were made on pristine and fluorinated bilayer graphene using standard electron-beam lithography (JEOL 9500FSZ at 100 keV) and e-beam evaporation. Back gate voltages were applied to the highly p-doped Si substrate with a 300-nm thick $SiO_2$ layer to make the field effect transistor structure.

**Fabrication $MoS_2$ device with FBLG insertion layer.** Multilayer $MoS_2$ was first mechanically exfoliated from bulk materials (HQ graphene) on an $SiO_2$/Si substrate. BLG was stacked on the $MoS_2$ flake by an aligned transfer under a microscope[34,35]. The sample then is spin coated with PMMA, and the contact regions are opened using electron-beam lithography. Plasma fluorination is applied to induce FBLG, and metals with different work functions including Ti, Cr, Au, and Pd with a thickness of 20 nm are deposited as the electrical contacts. To protect the metals from oxidation, all the contacts are covered by a thick Au layer (100 nm).

**Raman and XPS spectroscopies.** Raman spectra were acquired at room temperature with an excitation wavelength of 532 nm. XPS measurements were conducted at room temperature under a high vacuum of $10^{-9}$ mbar.

**Electrical characterization.** Electrical transport measurement of the fabricated devices were performed using a semiconductor parameter analyzer (Keithley 2000) in a vacuum probe-station which can heat the sample state at different temperatures.

## ASSOCIATED CONTENT

**Supporting Information.**

Supporting information includes Figures S1-S5 showing the Raman spectra of pristine and fluorinated BLG, and the transfer curve of the $MoS_2$ transistors fabricated with different metal contacts without and with FBLG and the extracted SBH at the metal/$MoS_2$ interfaces.

## AUTHOR INFORMATION

**Corresponding Author**

   **Manh-Ha Doan**; E-mail address: mando@dtu.dk

   **Peter Bøggild**; E-mail address: pbog@dtu.dk

**Notes**

The authors declare no competing financial interest.

## ACKNOWLEDGMENTS

This work was supported by the Danish National Research Foundation (DNRF) Center for Nanostructured Graphene (DNRF103) and EU Graphene Flagship Core 3 (881603). Cheng Xiang acknowledges the support from the China Scholarship Council (Grant No. 201906240046).

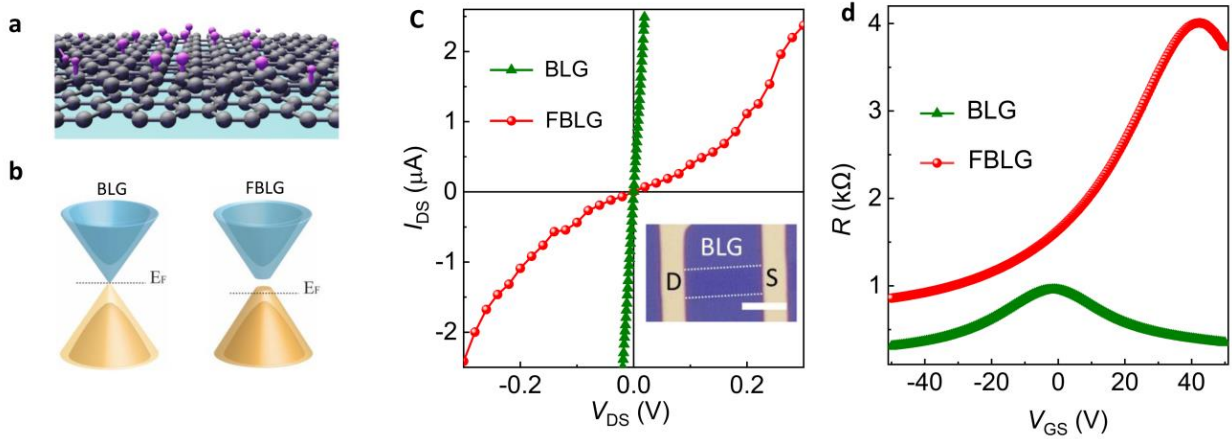

**Figure 1.** (**a**) Schematic of a cross-sectional view of FBLG on the substrate. Gray circles denote the carbon (C) atoms and the pink ones represent the fluoride (F) atoms. Covalent C-F bondings form at the surface of the top layer of BLG after fluorination. (**b**) Band diagram of pristine BLG and FBLG. (**c**) Current-voltage characteristics of BLG and FBLG at $V_G = 0$ V. The inset shows an optical microscopy image of the device. Scale bar is 5 µm. (**d**) Dependence of the resistivity of the BLG and FBLG on the back-gate voltage. Higher on-off ratio observed in the FBLG device suggests the presence of a band gap opening after fluorination.

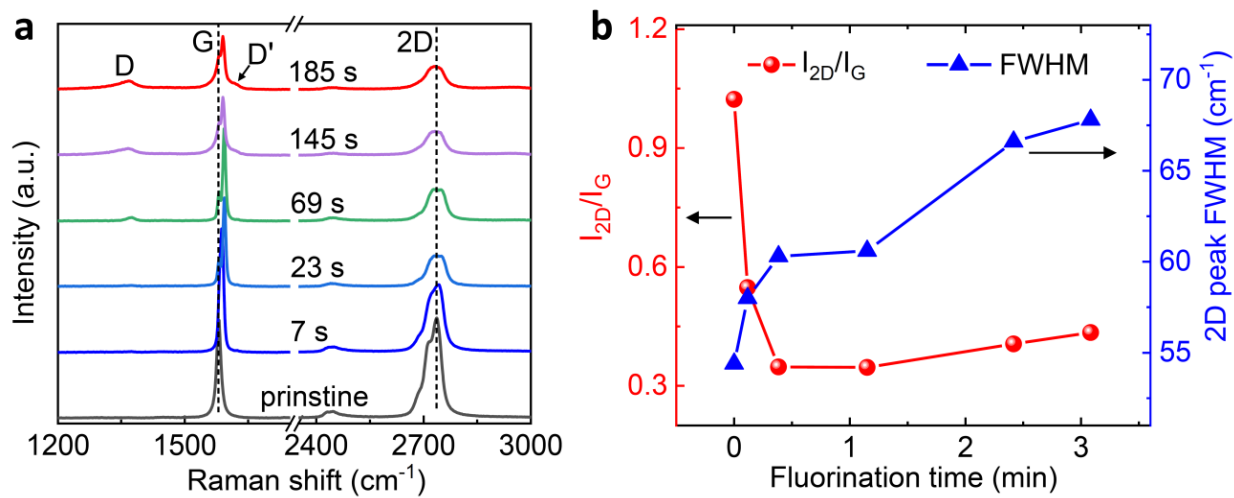

**Figure 2.** (**a**) Evolution of the Raman spectra of BLG with increasing fluorination time. (**b**) Intensity and full width of half-maximum (FWHM) of the 2D peak as a function of fluorination time.

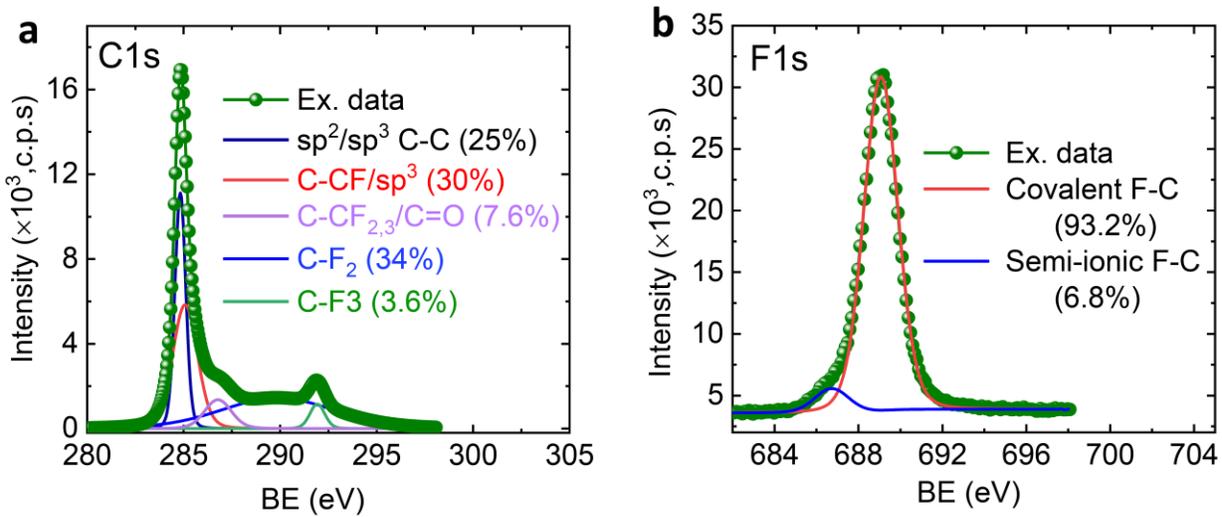

**Figure 3.** XPS spectra of FBLG in the (**a**) C1s and (**b**) F1s ranges.

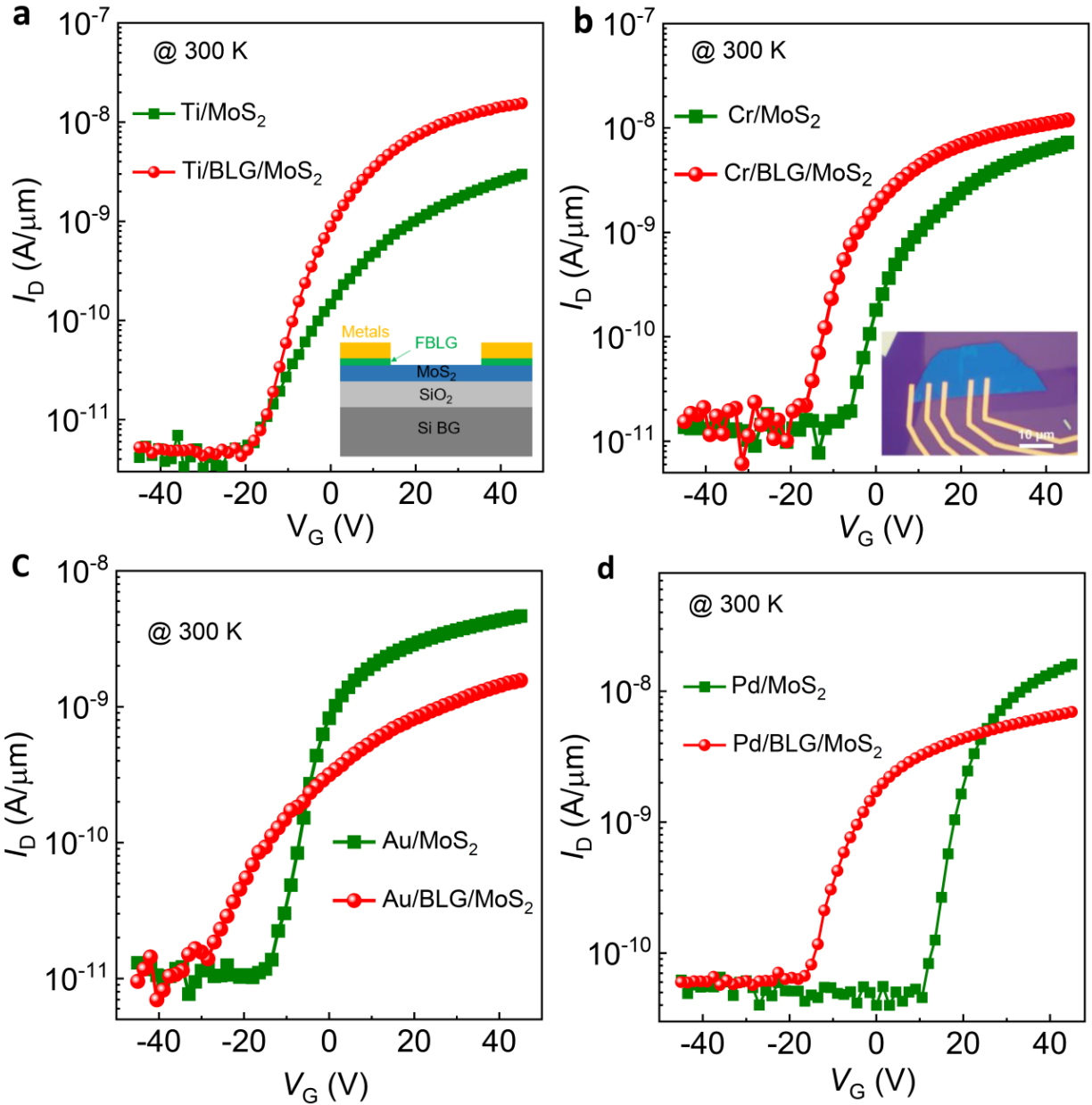

**Figure 4.** Transfer curves of the MoS$_2$ transistors without and with FBLG insertion into the metal/MoS$_2$ interfaces (**a**-**d**). The insets of Figure 4a and 4b show a cross-sectional schematic and the optical microscopy image of MoS$_2$ device.

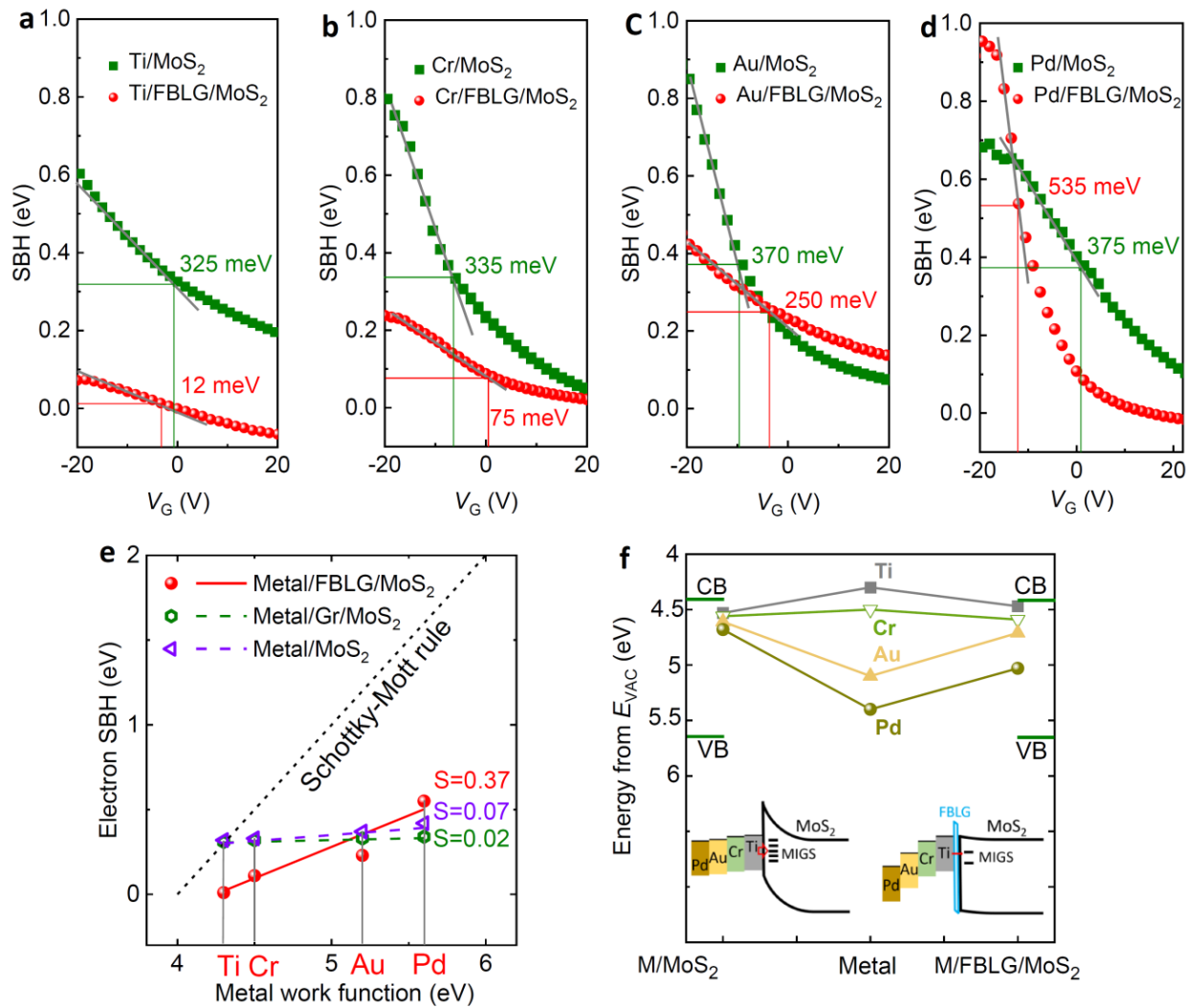

**Figure 5.** (**a-d**) Effective electron Schottky barrier height of the metal/MoS$_2$ contacts without and with FBLG. (**e**) Extracted pinning factor (S) for the MoS$_2$ devices with different contacts. (**f**) Alignment of the Fermi level in MoS$_2$ with the metal contacts without and with inserted FBLG.